\documentclass[12pt,a4paper]{article}
\usepackage{amsmath}
\usepackage{amsfonts}
\usepackage{amssymb}
\usepackage{graphicx}
\usepackage{geometry}
\usepackage{dcolumn}
\usepackage{float}
\usepackage{multirow}
\usepackage{tikz}
\usepackage{authblk}
\usepackage[hidelinks]{hyperref} 
\hypersetup{
    colorlinks=true,    
    urlcolor=blue,
    citecolor=blue,
    linkcolor=blue
    }
\usepackage{algorithmic}
\usepackage{algorithm}
\usepackage{amsthm}

\floatname{algorithm}{Periodización histórica por mayoría simple} 
\usepackage[autostyle]{csquotes} 
\usepackage{setspace} 
\usepackage{color} 
\hypersetup{pdftitle={¿Cuánto es demasiada inflación?}}
\usepackage[hyperref=true,backend=biber,style=apa,sorting=nyt]{biblatex}
\DefineBibliographyStrings{spanish}{andothers = {et\addabbrvspace al\adddot}} 
\usepackage[spanish, es-tabla]{babel}
\usepackage[sc]{mathpazo}
\usepackage[T1]{fontenc}

\begin{document}

\title{\textbf{¿Cuánto es demasiada inflación? 
\\
Una clasificación de regímenes inflacionarios}}

\author[1]{Manuel de Mier\thanks{Email: manueldemier@gmail.com. Agradecemos a Fernando Tohmé por sus valiosos comentarios a distintas versiones de este documento. Todo error u omisión es responsabilidad exclusiva de los autores. Este proyecto fue financiado por la beca Estímulo a las Vocaciones Científicas del Consejo Interuniversitario Nacional.}}
\author[1,2]{Fernando Delbianco}

\affil[1]{Departmento de Economía, Universidad Nacional del Sur}
\affil[2]{INMABB-CONICET, Bahía Blanca, Argentina}


\date{Noviembre 2023}

\maketitle
\spacing{1.3}
\begin{abstract}
Las clasificaciones de regímenes inflacionarios propuestas en la literatura se han basado en su mayoría en caracterizaciones arbitrarias, sujetas a juicios de valor por parte de los investigadores. El objetivo de este estudio es proponer un nuevo enfoque metodológico que reduzca la subjetividad y mejore la precisión en la construcción de dichos regímenes. El método se construye a partir de una combinación de técnicas de \textit{clustering} y árboles de clasificación, que permite obtener una periodización histórica de la historia inflacionaria argentina para el período 1943-2022. Adicionalmente, se introducen dos procedimientos para suavizar la clasificación en el tiempo: una medida de contigüidad temporal de las observaciones y un método de suavización móvil basado en la regla de mayoría simple. Los regímenes obtenidos son contrastados con la literatura existente para la relación inflación-precios relativos, encontrando un mejor desempeño de los regímenes propuestos. \medskip
\\
\textbf{Palabras Clave:} Inflación, Periodización histórica, \textit{Clustering}, Árboles de clasificación.
\\
\textbf{Códigos JEL:} C38, E31, N16.
\end{abstract}

\clearpage
\spacing{1.5}
\section{Introducción}

A pesar de ser la inflación uno de los temas más relevantes de la ciencia económica, diversos aspectos en torno a su caracterización han sido vagamente explorados por la literatura. En particular, la noción de ``régimen inflacionario'', a pesar de su extenso uso, se fundamenta en criterios subjetivos sin un consenso existente entre los investigadores. Esto resulta problemático dado que no es independiente de juicios de valor y está potencialmente sujeto a sesgos y errores de clasificación.

Se han planteado diferentes alternativas en la literatura acerca de la cantidad de regímenes inflacionarios y los límites que los definen, sin arribar a un consenso. En general, estas caracterizaciones se asocian a la definición de umbrales numéricos (\cite{bruno1998}; \cite{cagan1956}; \cite{dabus2000}; \cite{DSW1990}; \cite{dornbuschfisher1993}; \cite{werner2000}), con excepción de \textcite{HL1995}, quienes se basan en conductas de comportamiento de los agentes. Además, debido a su construcción general, estas clasificaciones no capturan las heterogeneidades inherentes a los procesos inflacionarios de cada país, sino que generalizan un conjunto de regímenes inflacionarios para todos ellos.

El objetivo de este trabajo consiste en realizar una caracterización, clasificación y periodización de los regímenes inflacionarios para Argentina, que mejore en términos metodológicos con respecto a la literatura existente, mediante el uso de técnicas de aprendizaje automático. Estos métodos permiten enriquecer el análisis de datos macroeconómicos, eliminando arbitrariedades en su proceso de estudio y capturando las particularidades de cada país.

El enfoque metodológico se fundamenta en los trabajos realizados por \textcite{fermercosur2021} y \textcite{LYS2005}, quienes emplean técnicas de \textit{clustering} para el estudio de la historia del MERCOSUR y la elaboración de regímenes cambiarios, respectivamente. En línea con estos estudios, se emplea el algoritmo de $k$-medias para la construcción de los regímenes inflacionarios, a los cuales luego se les calculan sus valores umbrales mediante el método basado en árboles de decisión CART. Posteriormente, los \textit{clusters} obtenidos son validados mediante el análisis ANOVA y el análisis de \textit{silhoutte}.

Debido a la naturaleza del problema bajo estudio, se requiere suavizar en el tiempo la clasificación obtenida, es por ello que se introduce una nueva medida denominada Distancia de Contigüidad Temporal (DCT), la cual modifica la distancia entre las observaciones según su proximidad temporal, mejorando la agrupación en el tiempo haciendo más probable que dos observaciones contiguas pertenezcan al mismo \textit{cluster}. Como alternativa, también se emplea la regla de votación de mayoría simple dentro de una ventana temporal móvil, para asignar el régimen más frecuente de la misma a cada observación. Por último, se estudia el desempeño de los distintos regímenes propuestos para la relación inflación-precios relativos, en comparación con el análisis realizado por \textcite{dabus2000} y \textcite{caraballoregim}.

Este trabajo realiza diversas contribuciones. En primer lugar, se presenta una caracterización de regímenes inflacionarios específica para la historia inflacionaria argentina, la cual por su construcción resulta menos arbitraria que las presentes en estudios previos. A su vez, se introducen dos procedimientos que permiten una periodización más suave en el tiempo, facilitando la interpretación y el análisis histórico. Por último, en términos de poder explicativo, las clasificaciones propuestas mostraron un mejor desempeño en relación a las provistas por la literatura.

El resto del trabajo se estructura de la siguiente manera. La sección 2 desarrolla una revisión de la literatura sobre las distintas definiciones y características de los regímenes inflacionarios. En la sección 3 se aborda el enfoque metodológico a adoptar y las fuentes de información. La sección 4 presenta las estimaciones de los regímenes, la periodización histórica de la inflación argentina y finaliza con la comparación con estudios previos para la relación inflación-precios relativos. Por último, la sección 5 expone las conclusiones del trabajo.

\section{Revisión de la literatura}

Existen diversos criterios para clasificar a la inflación, uno de los más empleados consiste en categorizarla en distintos regímenes en base a la tasa a la cual se incrementan los precios para una cierta unidad temporal \parencite{frisch1984}.

La investigación de \textcite{cagan1956} sobre las dinámicas de la hiperinflación constituye uno de los primeros antecedentes en la temática. En la misma, se delimita el inicio de una hiperinflación cuando el aumento de precios es de, por lo menos, un 50\% por mes (12.875\% anual), y su fin cuando se mantiene sostenidamente por debajo de dicho umbral durante al menos un año.

Sin embargo, diversos autores plantearon definiciones alternativas para la hiperinflación. Por ejemplo, \textcite{DSW1990} en su estudio sobre episodios inflacionarios extremos rechazan la regla de Cagan, utilizando un umbral menor para la hiperinflación de 1.000\% anual (implicando tasas mensuales por encima de 15 a 20\%, sostenidas por varios meses).

Para \textcite{kiguel1989}, la hiperinflación es vista como un proceso intrínsecamente inestable, el cual los países pueden experimentar incluso a niveles menores al 50\% impuesto por Cagan. En tal sentido, \textcite{dornbusch1992} señala que la distinción entre la hiperinflación y casos de inflación extrema pero levemente menor es arbitraria, debido a que el comportamiento de la economía entre estos niveles es virtualmente idéntico con la inflación como el factor dominante de la economía. Bajo esta concepción, en un estudio posterior, \textcite{kiguel1995} flexibilizan la regla de Cagan para delimitar correctamente el fin de la hiperinflación peruana de 1988-1990, y así poder analizar la dinámica no explosiva que presentó dicho episodio. Además, los autores caracterizan un régimen de alta inflación a tasas entre 20\% y 49\% mensual (792-11.874\% anual).

Un criterio alternativo es provisto por \textcite{bruno1998}, quienes en su estudio de crisis de alta inflación consideran un umbral de 40\% anual durante al menos dos años consecutivos para clasificar dichos episodios. Este límite es determinado en base a la observación de un quiebre en la dinámica económica cercano a dicho valor, a partir del cual la inflación tiende a acelerarse y volverse más volátil, aumentando la probabilidad de conducir a una hiperinflación.

En contraposición a estos estudios, se encuentra la caracterización brindada por \textcite{HL1995}, quienes plantean los regímenes inflacionarios en términos del horizonte de planeamiento de los agentes, expresado en qué unidad temporal se informa la tasa de inflación. De esta manera, la inflación es moderada cuando los individuos la expresan en términos anuales, alta cuando este horizonte pasa a ser de un mes, e hiperinflación cuando es menor al mes. Los autores proporcionan un rango aproximado de 5\% a 50\% mensual para el régimen de alta inflación, mientras que acuerdan con \textcite{cagan1956} para la hiperinflación. Se puede notar la gran amplitud del régimen de alta inflación, con una tasa anual equivalente que va desde 80\% a 12.875\% anual.

En relación a la inflación moderada, \textcite{brunner1973} la caracterizan como un estado en el cual los precios crecen a tasas anuales raramente por encima del 10\%, con desaceleraciones luego de picos inflacionarios relativamente altos y con la posibilidad de algunos episodios deflacionarios. En línea con esto, \textcite{werner2000} estudian las dinámicas y transiciones del régimen de inflación moderada (10-20\% anual), con respecto al de inflación baja (<10\%) y al de inflación alta (>20\%). Por otro lado, \textcite{dornbuschfisher1993} la especifican en tasas entre 15\% y 30\% anual, por al menos tres años consecutivos. Finalmente, \textcite{dabus2000} en su estudio para la Argentina, caracteriza la inflación moderada cuando la inflación mensual se encuentra por debajo del 2\%, alta entre 2\% y 10\%, muy alta entre 10\% y 50\%, e hiperinflación para valores mayores al 50\%, lo cual la hace consistente con la regla de Cagan. 

Como puede notarse, no existe un consenso en la literatura acerca de la cantidad de regímenes ni de los límites que los definen. Esto es producto de la arbitrariedad que poseen estas definiciones, la cual ya era reconocida por varios de sus autores (\cite{bruno1998}; \cite{cagan1956}; \cite{dornbusch1992}; \cite{werner2000}). A excepción de \textcite{bruno1998}, las caracterizaciones provistas no están basadas en criterios objetivos o métodos estadísticos, sino que se fundamentan en los propios procesos inflacionarios que se deseaban analizar, en juicios de valor o en convenciones que la ciencia económica adoptó con el paso del tiempo, como la regla de Cagan.

Recientemente, diversos autores emplearon modelos autorregresivos de umbrales para la determinación de regímenes. En estos trabajos los regímenes son definidos en base a la búsqueda de quiebres en las relaciones económicas de interés, como la relación entre inflación y precios relativos \parencite{bicknautz2008}, o la inflación y el \textit{pass-through} cambiario (\cite{brufman2017}; \cite{cheikh2016}; \cite{costa2019}). La limitación de estos procedimientos es que los regímenes obtenidos no son independientes de la relación de interés que se evalúa. 

En línea con estos estudios, \textcite{castagninodamato2008es} y \textcite{damatogaregnani2013} emplean pruebas de cambio de estructura para identificar quiebres en un proceso autorregresivo de la inflación argentina, caracterizando los regímenes en función de los resultados de dichas pruebas. Sin embargo, es importante destacar que los autores conciben estos regímenes en un sentido más amplio de lo habitual, refiriéndose a regímenes monetarios que implican un cierto marco institucional de la política económica.

Un rasgo relevante que capturan los modelos autorregresivos es que la dinámica de la economía y de los precios es diferente bajo distintos niveles de inflación, producto de los cambios en el comportamiento de los agentes económicos y en los mecanismos de formación de precios. Diversos autores han estudiado la relación inflación-precios relativos, encontrando la presencia de quiebres en la dinámica entre las variables para distintos regímenes (\cite{caraballoregim}; \cite{dabus2000}; \cite{elias1996}). En general, la inflación más alta es más volátil \parencite{taylor1981}, y se asocia con una mayor dispersión de precios (\cite{moll}; \cite{tommasi1992}). Esto se debe a que los episodios de inflación moderada suelen ser más estables, reforzados por mecanismos como la indexación que dificultan su baja, mientras que la hiperinflación presenta una dinámica inestable y explosiva. En estos casos extremos, puede existir un deterioro significativo de la economía. En este sentido, \textcite{bruno1998} presentan evidencia de una relación negativa entre crecimiento económico e inflación para las crisis de alta inflación.


Además de las características propias de los distintos regímenes, estos también presentan variaciones sustanciales entre países. Las diferencias se deben a las distintas percepciones del público sobre la inflación, a los shocks que originaron el proceso, así como a la historia inflacionaria previa y otras características específicas de la estructura económica que afectan la dinámica de precios. Un ejemplo de esto es el efecto histéresis, que consiste en una persistente baja demanda por la moneda local provocado por la historia de alta inflación \parencite{DDF2016}.

Estas diferencias pueden observarse en los procesos de alta inflación e hiperinflación de América Latina durante la década de 1980, los cuales presentan diferencias sustanciales con las experiencias europeas de la década de 1920. La duración de los primeros fue más extensa, superando una década, con niveles de inflación relativamente menores y con un patrón de \textit{stop-and-go}, marcado por estabilizaciones temporales seguidas de estallidos inflacionarios \parencite{DSW1990}. Por otro lado, las hiperinflaciones europeas se originaron en factores externos como la guerra, mientras que en las latinoamericanas estos fueron secundarios, siendo los factores domésticos los causantes principales \parencite{dornbusch1992}. Según \textcite{kiguel1995}, estas diferencias se deben a la prolongada historia de alta inflación no explosiva de las experiencias latinoamericanas, que provocó que pequeños shocks económicos logren desestabilizar las economías y conducirlas hacia la hiperinflación.

En particular, la hiperinflación argentina presentó similitudes en cuanto a su duración e intensidad con las experiencias europeas, mientras que en el ámbito fiscal se observaron importantes discrepancias. El mayor control gubernamental y los mecanismos de indexación, resultantes de la prolongada historia inflacionaria argentina permitieron evitar el colapso de los ingresos fiscales como sucedió en las experiencias europeas \parencite{kiguel1995}. En estas últimas, el origen de la hiperinflación se debió a un fuerte aumento repentino del déficit fiscal, mientras que en Argentina se originó a partir de un prolongado proceso de deterioro de las cuentas públicas.

Para la inflación argentina también es posible identificar rasgos distintivos que la separan de las demás experiencias de América Latina. La fuerte influencia de la inercia inflacionaria es una característica que la diferencia del resto de la región. En particular, la persistencia de la inflación es el factor que explica en mayor medida la variabilidad de la misma, representando un 80\% de la volatilidad inflacionaria para el período 2004-2019 \parencite{cicco2022}.

Estos rasgos distintivos de los regímenes inflacionarios pueden sintetizarse en la siguiente definición. Siguiendo a \textcite{tohme2005}, se define un régimen inflacionario como un contexto económico caracterizado por una tasa de inflación fluctuando en un cierto rango de valores, a los cuales se les asocia un cierto grado de incertidumbre y un sistema de expectativas. Esta caracterización entiende a cada régimen como un cierto patrón de conexiones de la economía, asociado a un estado específico del sistema económico. Cada uno de estos estados tienen como consecuencia episodios inflacionarios de distinta magnitud, en concordancia con las experiencias pasadas y los shocks de la política económica. Esta concepción permite capturar el hecho de que cada economía presentará regímenes diferentes, y que cada una de ellas funcionará de manera distinta dependiendo en qué régimen se encuentra y por cuánto tiempo lo hace.

De esta revisión surge la necesidad de mejorar los métodos para identificar los umbrales de los diferentes regímenes inflacionarios, teniendo en cuenta la heterogeneidad de los países y las características propias de cada régimen. Aquí es donde toma relevancia la utilización de métodos de aprendizaje automático que van a permitir, bajo mínimas premisas, caracterizar, clasificar y periodizar de un modo óptimo a la inflación en distintos regímenes inflacionarios particulares para la Argentina.

\section{Metodología y fuentes de información}

\subsection{Método de agrupamiento y clasificación}

Debido a que no se parte de una clasificación previa de regímenes de la literatura, es necesario recurrir a un procedimiento que genere de manera endógena las diferentes categorías junto a sus respectivos umbrales. Un enfoque aplicado en la literatura de análisis simbólico consiste en seleccionar el umbral que minimice la entropía normalizada de Shannon \parencite{brida2011}. Otra alternativa es segmentar el espacio de variables mediante la división en subconjuntos de valores superiores e inferiores a las medias de cada una (\cite{london2011}; \cite{london2022}). Este último enfoque presenta limitaciones, ya que la media puede no ser un umbral representativo de las dinámicas entre regímenes. 

En \textcite{caraballoregim}, se desarrolla una clasificación de regímenes inflacionarios en dos etapas. En la primera, se realiza una periodización mediante la búsqueda de quiebres en una serie suavizada de la inflación. En la segunda etapa, se asigna a cada período determinado un régimen en función de su tasa de inflación promedio, utilizando los regímenes presentados en \textcite{dabus2000}. Sin embargo, este enfoque posee ciertas desventajas, como la necesidad de una clasificación previa de regímenes inflacionarios y la falta de detección de transiciones suaves entre regímenes.

En este trabajo, se optó por la utilización de técnicas de \textit{clustering}, comúnmente utilizadas para identificar patrones y estructuras en los datos. La idea detrás de estos métodos es agrupar a las observaciones en grupos denominados \textit{``clusters''}, que poseen la característica de estar compuestos por unidades que son lo más similares entre sí, en relación a las pertenecientes a los demás \textit{clusters}. La similitud entre los datos se define utilizando una métrica calculada sobre las observaciones, como por ejemplo, la distancia euclídea. Estos grupos generados son posteriormente interpretados como regímenes inflacionarios.

Específicamente, se emplea el algoritmo de $k$-medias, el cual particiona el conjunto de observaciones en $k$ \textit{clusters}, con cada observación perteneciendo a un único \textit{cluster}. Este algoritmo obtiene iterativamente una partición óptima mediante la minimización de las distancias euclídeas entre las observaciones de un mismo \textit{cluster} y su centro, definido como la media de las observaciones pertenecientes a dicho \textit{cluster} \parencite{kasam2017}. 

Este método sobresale en este tipo de aplicaciones en comparación con otras técnicas de \textit{clustering}, como los métodos jerárquicos, debido a que requiere una intervención mínima por parte del investigador en cuanto a los criterios de clasificación utilizados \parencite{LYS2005}. Adicionalmente, su sensibilidad a \textit{outliers}, que en otros contextos resulta problemática, aquí se destaca como una ventaja ya que permite identificar y aislar correctamente el régimen hiperinflacionario. Alternativas como $k$-medianas o los métodos jerárquicos presentan dificultades en este sentido, asignando erróneamente observaciones en los \textit{clusters}. Por otro lado, $k$-medias ha sido la empleada por la literatura en contextos similares, donde se buscaba reducir la arbitrariedad de clasificaciones existentes. \textcite{fermercosur2021} emplean este método para obtener una periodización histórica del MERCOSUR basada en índices de comercio exterior, mientras que \textcite{LYS2005} lo utilizan para la construcción de una clasificación de regímenes cambiarios \textit{de facto} en base al comportamiento de diversas variables del mercado cambiario.

A pesar de que existen criterios para determinar la cantidad óptima de \textit{clusters}, estos se ven perturbados por la presencia de \textit{outliers} en la muestra, como los episodios hiperinflacionarios. En consecuencia, todos los indicadores sugieren en favor de la construcción de dos grupos que logren separar dichas observaciones del resto de la muestra (véase Apéndice \ref{appendix:appA}), lo cual no permitiría capturar los regímenes que existen para menores niveles de inflación. Además, la no correcta separabilidad de los \textit{clusters} y las diferencias significativas en los tamaños de las muestras influyen en que criterios como el \textit{gap statistic} no funcionen adecuadamente (\cite{tibshirani2001}; \cite{yin2008}). Por estos motivos, se predefinió identificar cuatro regímenes inflacionarios: inflación baja, inflación moderada, inflación alta e hiperinflación. 

Para determinar los umbrales de los regímenes generados se emplea el método de clasificación basado en árboles CART \parencite{breiman}. Esta técnica utiliza los grupos obtenidos por $k$-medias como insumo, y construye un árbol de decisión que busca predecirlos, particionando recursivamente el espacio de las variables mediante valores óptimos de corte \parencite{murphyML}. Estos límites definidos en términos de las variables constituyen los umbrales de los regímenes inflacionarios. Por último, con el fin de establecer si se realizó un agrupamiento forzoso de los datos, se validan internamente los \textit{clusters} obtenidos mediante el análisis de varianza (ANOVA) y el análisis de \textit{silhoutte}.

\subsection{Periodización histórica} \label{cap:section_per}

Para lograr una periodización histórica efectiva, es fundamental que los regímenes asignados exhiban cierta suavidad a lo largo del tiempo, esto implica que no existan cambios abruptos de régimen, que se reviertan en cortos períodos de tiempo. En estudios previos, el uso de datos anuales al realizar el análisis de \textit{clusters} permitía aminorar esta problemática. En cambio, en el presente trabajo la incorporación de datos mensuales introduce una complejidad adicional en este aspecto. Hasta la fecha, a pesar de existir técnicas de \textit{clustering} que toman en consideración la contigüidad espacial de las observaciones para lograr un mejor agrupamiento \parencite{quintana2021}, no se han desarrollado métodos que consideren la contigüidad temporal para agrupar observaciones de series de tiempo.

Para abordar esta limitación se presenta una nueva medida, llamada Distancia de Contigüidad Temporal (DCT)\footnote{ Estrictamente, la DCT constituye una medida de disimilitud, dado que cumple las condiciones de no negatividad, identidad y simetría, pero no la de desigualdad triangular. De todas formas, ninguna de estas propiedades son esenciales para realizar un análisis de \textit{clusters} \parencite{kaufman}.}, basada en la reponderación de la distancia euclídea entre las observaciones, según su cercanía temporal. La DCT permite suavizar la clusterización en el tiempo, ya que aumenta las distancias originales a medida que las observaciones se alejan temporalmente, reduciendo la probabilidad de que pertenezcan al mismo \textit{cluster}. Como ejemplo, para una serie temporal univariada, dos observaciones consecutivas reciben un aumento mínimo en su distancia, mientras que para la primera y última observación de la serie el aumento es el máximo posible.

Formalmente, considere una matriz $X$ de dimensión ($T \times p$), con $p$ variables seguidas a lo largo de $T$ períodos en el tiempo ($t=1,2,...\,,T$), y sea $\lambda$ un número real no negativo. Para un $\lambda$ dado, la DCT entre dos observaciones $t$ y $s$ es definida como:

\begin{equation} \label{eqn:eq1}
\displaystyle DCT(t,s) =\left( 1+\lambda \frac{|t-s|}{T-1}\right)\sqrt{\sum\limits _{k=1}^{p}( X_{tk} -X_{sk})^{2}}  \ \ \ \ \ \ \forall \ t,s \in T
\end{equation}

De la ecuación \eqref{eqn:eq1} se puede observar que el parámetro $\lambda$ regula el suavizado temporal, ya que cuando el valor de $\lambda$ es igual a cero, se obtiene la distancia euclídea original, en tanto que para valores mayores de $\lambda$, las distancias euclídeas crecen como máximo por un factor de $1+\lambda$. Además, al dividir por la máxima distancia posible $T-1$ se relativiza el factor a la escala temporal de los datos.

Debido a que el algoritmo de $k$-medias no puede emplearse con una matriz de disimilitudes, es necesario transformar la matriz de DCT en una representación en el espacio de las $p$ variables. Esto se realiza mediante el análisis de coordenadas principales no métricas, el cual brinda una representación óptima que mejor ajusta a la matriz de disimilitudes. Posteriormente, se aplica el método de $k$-medias a esta matriz transformada. Esta estrategia se sustenta en \textcite{deluca2011}, quienes demuestran mediante simulaciones que este procedimiento en dos etapas presenta un rendimiento superior a la aplicación de métodos jerárquicos, los cuales pueden aplicarse directamente sobre la matriz de disimilitudes.

Una estrategia alternativa para suavizar la clasificación en el tiempo implica emplear una regla de votación, que consiste en considerar a cada observación como un votante que emite un voto en base a su régimen asociado. Esta regla se aplica en una ventana temporal móvil sobre la serie de regímenes, es decir, sobre la serie que posee para cada momento en el tiempo el régimen inflacionario que generó el análisis de \textit{clusters}. En este caso, se opta por utilizar la regla de mayoría simple. El procedimiento construido toma una ventana temporal de la observación $t$ hasta $h$ rezagos, dentro de la cual se cuentan los distintos regímenes asignados. Por ende, el intervalo de votación posee $h+1$ observaciones. De existir un régimen que cumpla la regla de mayoría simple, este se le asigna a la observación $t$. Cabe aclarar que en todo momento se vota en base a los datos originales, sin tener en cuenta las votaciones previas. Una presentación más detallada del procedimiento se expone en el siguiente esquema: 

\vspace{3mm}
\begin{algorithm}[H]
\caption{}
\begin{enumerate}

\item Sea $X_t$ la serie de regímenes y $h \in 	\mathbb{N}$ la cantidad de rezagos considerados.
\item Para $t=1,\,2,...\,,T$:
\begin{enumerate}
\item Si $t\leq h$, mantener el régimen original de la observación $t$.
\item Si $t>h$, contar la cantidad de observaciones $k_c$ del régimen $C$ en el intervalo $[t-h,t]$. Si existe una clase $C^*$ tal que $k_{C^*}\!\!>\!\frac{h+1}{2}$, dicho régimen es considerado ganador por mayoría simple y es asignado a la observación $t$. Caso contrario, se mantiene el régimen original.

\end{enumerate}
\end{enumerate}
\end{algorithm}

A modo de ejemplo, la Tabla \ref{tab:ejemploms} ilustra la periodización para una serie de 10 datos y una ventana temporal con 2 rezagos.

\begin{table}[H]
\centering
\caption{Ejemplo periodización por mayoría simple}
\footnotesize
\vspace{2mm}
\label{tab:ejemploms}
\begin{tabular}{ccc}
\hline
$t$ & $X_t$ & $X_t$ (Mayoría simple) \\ 
\hline
1 & Inflación baja & Inflación baja \\ 
2 & Inflación moderada & Inflación moderada \\ 
3 & Inflación moderada & Inflación moderada \\
4 & Inflación baja & Inflación moderada \\
5 & Inflación moderada & Inflación moderada \\
6 & Inflación alta & Inflación alta \\ 
7 & Hiperinflación & Hiperinflación \\ 
8 & Inflación moderada & Inflación moderada \\
9 & Inflación baja & Inflación baja \\
10 & Inflación moderada & Inflación moderada \\
\hline
\end{tabular}
\end{table}

De esta manera, las primeras 2 observaciones de la serie se mantienen inalteradas, así como aquellas en las que no resulta un régimen ganador por mayoría simple ($t$= 6, 7, 8, 9). Esta forma de proceder garantiza que, por un lado, se suavice la serie en períodos de relativa estabilidad donde ocasionalmente puede haberse identificado un cambio de régimen por un shock transitorio, por ejemplo, un incremento puntual en las tarifas de los servicios públicos. Por otro lado, no se ven afectados los resultados en intervalos de alta inestabilidad económica caracterizados por fuertes cambios de régimen.

\subsection{Fuentes de información}

La falta de estadísticas históricas de precios a nivel capítulo, así como la manipulación de las estadísticas oficiales de inflación, plantean una restricción significativa para el análisis. En consecuencia, solo es posible utilizar la tasa de inflación mensual como variable de clasificación, no pudiendo considerar, por ejemplo, medidas de precios relativos o de expectativas que permitieran capturar la dinámica diferencial de cada régimen. Se consideró al período comprendido entre febrero de 1943 y diciembre de 2022, inclusive, utilizando los datos de inflación del Instituto Nacional de Estadística y Censos de la República Argentina (INDEC). Debido a la intervención de las estadísticas públicas en el período 2007-2015, se utiliza para dicha etapa los datos de inflación del sitio web \textit{inflacionverdadera.com} (\cite{cavallo2016}; \cite{benfordfer}).

\section{Resultados}

\subsection{Estimaciones de $k$-medias}

La Tabla \ref{tab:resk4} presenta los resultados de la estimación de $k$-medias para los cuatro regímenes definidos, junto con los umbrales obtenidos mediante CART, los cuales constituyen intervalos cerrados por su mínimo y abiertos por su máximo.

\begin{table}[H]
\begin{center}
\caption{Resultados $k$-medias}
\label{tab:resk4}
\vspace{2mm}
\footnotesize
\begin{tabular}{lcccc}
\hline
\multirow{2}{*}{Regímenes} & \multirow{2}{*}{Observaciones} &  \multirow{2}{*}{Inflación promedio} & \multicolumn{2}{c}{Umbrales} \\ 
 & &  & Mínimo & Máximo \\ 
\hline
Inflación baja & 738 & 1,36 & - & 5,02 \\ 
Inflación moderada & 175 & 8,66 & 5,02 & 16,60 \\ 
Inflación alta & 41 & 24,30 & 16,60 & 70,02 \\
Hiperinflación & 5 & 112,86 & 70,02 & - \\
\hline
\multicolumn{5}{c}{\footnotesize Nota: Las cifras corresponden a tasas de inflación mensual.}\\
\end{tabular}
\end{center}
\vspace{-0.5cm}
\end{table}

Los resultados exhiben una gran preponderancia del régimen de inflación baja, el cual constituye más del 75\% de la muestra total de meses. Es importante resaltar que los umbrales obtenidos exhiben valores elevados en relación a los presentados en la revisión de la literatura. Por ejemplo, en contraste con la definición propuesta por \textcite{cagan1956} que considera la hiperinflación a partir del 50\% mensual, en este estudio se establece que la transición hacia dicho régimen ocurre a partir de un umbral de 70,02\% mensual.

Para evaluar si los regímenes obtenidos surgieron de un agrupamiento forzoso de los datos, se realizó un análisis ANOVA, que se presenta en la Tabla \ref{tab:anova_k4}. Los resultados permiten concluir que existen diferencias significativas entre las medias de los regímenes, lo que sugiere que al menos un grupo difiere del resto. Para evaluar en detalle este último punto, se realizan dos pruebas de diferencias de medias\footnote{ Las letras representan las diferencias significativas entre grupos y permiten compararlos. Si todos los grupos tienen letras distintas, esto implica que hay diferencias estadísticamente significativas entre ellos.} entre pares de regímenes: la prueba de Diferencia Mínima Significativa de Fisher y la prueba de Diferencia Honestamente Significativa de Tukey, ambas ampliamente empleadas en el análisis ANOVA \parencite{kutneranova}. La prueba de Fisher permite concluir con un nivel de significatividad individual, es decir, permite concluir con un nivel $\alpha$ que dos regímenes difieren entre sí. Mientras que en la prueba de Tukey el nivel de significatividad es global, lo que implica que se considera una tasa de error global $\alpha$ al realizar las comparaciones múltiples. De ambas pruebas se concluye que existen diferencias significativas entre los regímenes obtenidos, indicando en favor de que no existe un agrupamiento forzoso de los datos y que los \textit{clusters} generados poseen una buena separabilidad.

\begin{table}[H]
\begin{center}
\caption{Análisis ANOVA}
\label{tab:anova_k4}
\vspace{2mm}
\footnotesize
\begin{tabular}{l>{\centering}p{1.7cm}>{\centering}p{1.7cm}>{\centering}p{1.5cm}cc}
  \hline
 & Grados de libertad & Suma de cuadrados & Cuadrado medio & \multirow{2}{*}{F} & \multirow{2}{*}{p-valor} \\ 
  \hline
Regímenes          & 3 & 8.50 & 2.834 & 1667.40 & 0.000 \\ 
Residual   & 955 & 1.62 & 0.002 &  &  \\ 
   \hline
\multicolumn{6}{l}{}\\[-0.5em]
\end{tabular}
\end{center}
\vspace{-0.5cm}
\end{table}

\begin{table}[H]
\begin{center}
\footnotesize
\begin{tabular}{lcc}
\multicolumn{3}{l}{\small Pruebas de diferencia de medias ($\alpha$=0,01)}\\
\multicolumn{3}{c}{}\\[-1.3em]
\hline
Regímenes & Fisher & Tukey \\ 
\hline
Inflación baja & a & a \\ 
Inflación moderada & b & b \\ 
Inflación alta & c & c \\
Hiperinflación & d & d \\
\hline
\end{tabular}
\end{center}
\vspace{-0.5cm}
\end{table}

\begin{figure}[H]
\centering
\caption{Análisis de \textit{Silhoutte}}
\label{fig:silhoutte}
\vspace{1mm}
\includegraphics[width=\textwidth]{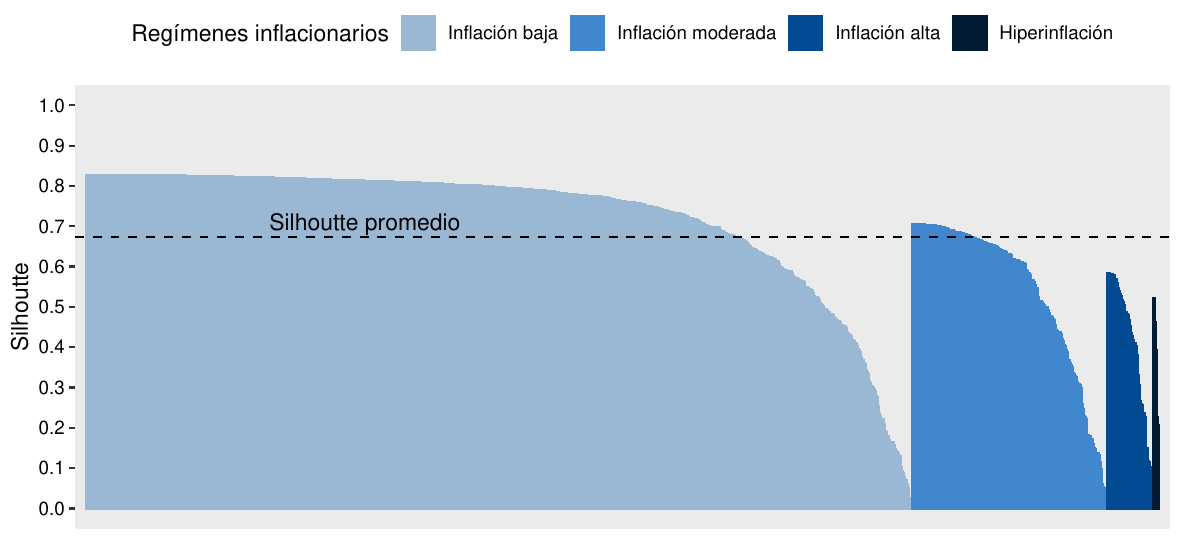}
\vspace{-9mm}
\end{figure}

Otra forma comúnmente empleada para evaluar la calidad de agrupamiento es el análisis de \textit{silhoutte}, el cual permite relacionar la cohesión dentro de cada \textit{cluster} con la separabilidad entre ellos \parencite{kasam2017}. Para cada observación se calcula un valor de \textit{silhoutte}, el cual mide qué tan cercano esta la observación con respecto a las de los \textit{clusters} vecinos. Un valor de \textit{silhoutte} cercano a 1 indica una correcta asignación de la observación al \textit{cluster}, mientras que valores negativos indican la presencia de observaciones mal clusterizadas, es decir, asignadas incorrectamente de grupo. Al observar la Figura \ref{fig:silhoutte}, se puede notar que no hay valores negativos, lo que indica que no hay observaciones asignadas incorrectamente de régimen inflacionario. A su vez, la \textit{silhoutte} promedio posee un valor relativamente alto de 0,67, principalmente influenciado por la cohesión dentro del régimen de inflación baja. Por otro lado, en los regímenes superiores de inflación existe una menor cohesión, lo cual puede asociarse a que capturan rangos más amplios de valores.

\begin{figure}[H]
\centering
\caption{Clasificación temporal ($k$-medias)}
\label{fig:cal_k4}
\vspace{1mm}
\makebox[\textwidth][c]{\includegraphics[width=1.1\textwidth]{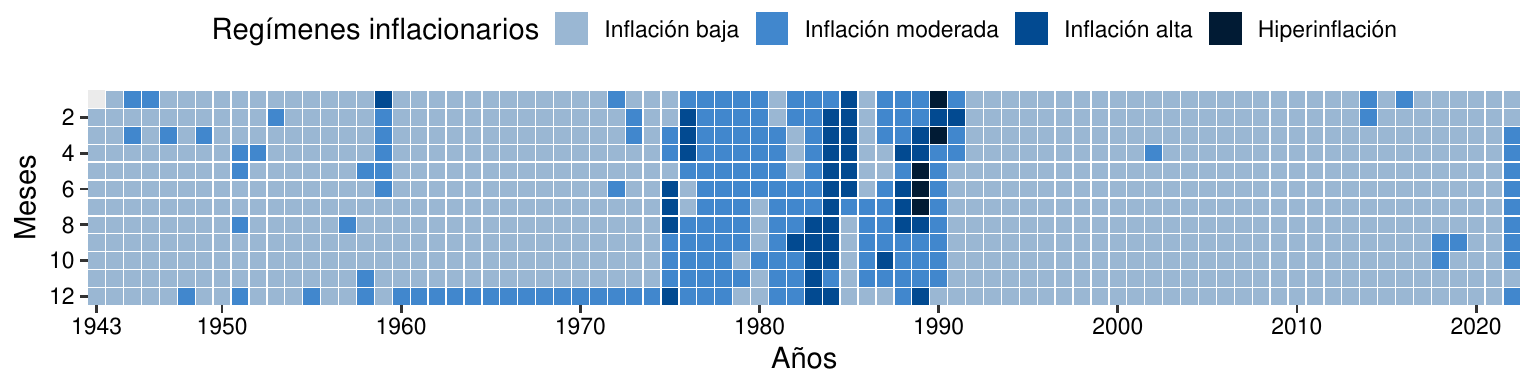}}
\vspace{-9mm}
\end{figure}

La Figura \ref{fig:cal_k4} presenta la clasificación en el tiempo de los regímenes estimados, donde cada cuadrado representa un mes, con un color que varía según el régimen inflacionario asociado a dicho mes. Puede notarse como gran parte de las observaciones pertenecientes a los regímenes de inflación moderada, alta e hiperinflación se encuentran concentrados entre los años 1975 y 1991. El primero se asocia al brote inflacionario causado por el ``Rodrigazo'', mientras que el segundo se corresponde con la desinflación producto de la implementación del régimen de Convertibilidad. Estos resultados son consistentes con la evidencia presentada por \textcite{castagninodamato2008es} y \textcite{damatogaregnani2013} sobre el impacto permanente en la dinámica inflacionaria del ``Rodrigazo'', que indujo una gran inestabilidad a la demanda de dinero y a las expectativas de los agentes. También, puede identificarse un episodio de inflación alta a comienzos de 1959, producto de la crisis ocurrida durante la presidencia de Arturo Frondizi. Por otro lado, los últimos datos de la muestra exhiben una transición de la economía argentina en marzo de 2022 al régimen de inflación moderada.

En la figura puede notarse la problemática de la suavización de la clasificación descrita en la sección \ref{cap:section_per}. Se observan diversos meses que, por incrementos puntuales en la inflación, inducen un cambio de régimen. Un ejemplo de esta situación puede observarse en los meses de diciembre de la década de 1960. A pesar de que el resto de las observaciones de la década son consistentemente clasificadas en el régimen de inflación baja, los datos de diciembre son clasificados en el régimen de inflación moderada.

\subsection{Periodización histórica}

La aplicación de la Distancia de Contigüidad Temporal (DCT) se llevó a cabo utilizando un parámetro $\lambda$ de 0,1, lo cual implica que las distancias originales entre los datos se expandirán como máximo en un 10\%. Por otro lado, para la periodización por mayoría simple (MS) se emplearon cuatro rezagos, lo que corresponde a una ventana de votación de cinco observaciones consecutivas. La Tabla \ref{tab:comp_per} presenta la comparación de resultados de las dos técnicas de suavización con respecto a la estimación de $k$-medias. Se puede observar como ambos procedimientos lograron su objetivo de disminuir la cantidad de cambios de régimen, en comparación con los resultados originales. Sin embargo, se observa una mayor reducción en la realizada por el procedimiento por mayoría simple. Por otro lado, la tabla presenta los cambios totales realizados con respecto a $k$-medias, siendo el de mayoría simple el que mayores reemplazos realizó.

\begin{table}[H]
\begin{center}
\caption{Comparación respecto a $k$-medias}
\label{tab:comp_per}
\vspace{2mm}
\footnotesize
\begin{tabular}{lccc}
  \hline
Regímenes & $k$-medias & DCT & MS\\ 
  \hline
Inflación baja 		& 738 & 724 &  771 \\ 
Inflación moderada 	& 175 & 183 &  144 \\ 
Inflación alta 		& 41  &  47 &  38 \\ 
Hiperinflación 		& 5   &   5 &   6 \\ 
   \hline
Cambios de régimen 	 & 135   &   112 &   30 \\
Cambios r/$k$-medias & -   & 58  &  92 \\
\hline
\end{tabular}
\end{center}
\vspace{-0.5cm}
\end{table}

Los regímenes obtenidos por estos métodos conservan los resultados obtenidos originalmente por el análisis ANOVA. Por otro lado, los índices de validación interna carecen de sentido cuando se emplean medidas de disimilitud como la DCT, perdiendo su habitual interpretación \parencite{deluca2021}.

\begin{figure}[H]
\centering
\caption{Clasificación temporal ($k$-medias, DCT, MS)}
\label{fig:comp_per}
\vspace{1mm}
\makebox[\textwidth][c]{\includegraphics[width=1.1\textwidth]{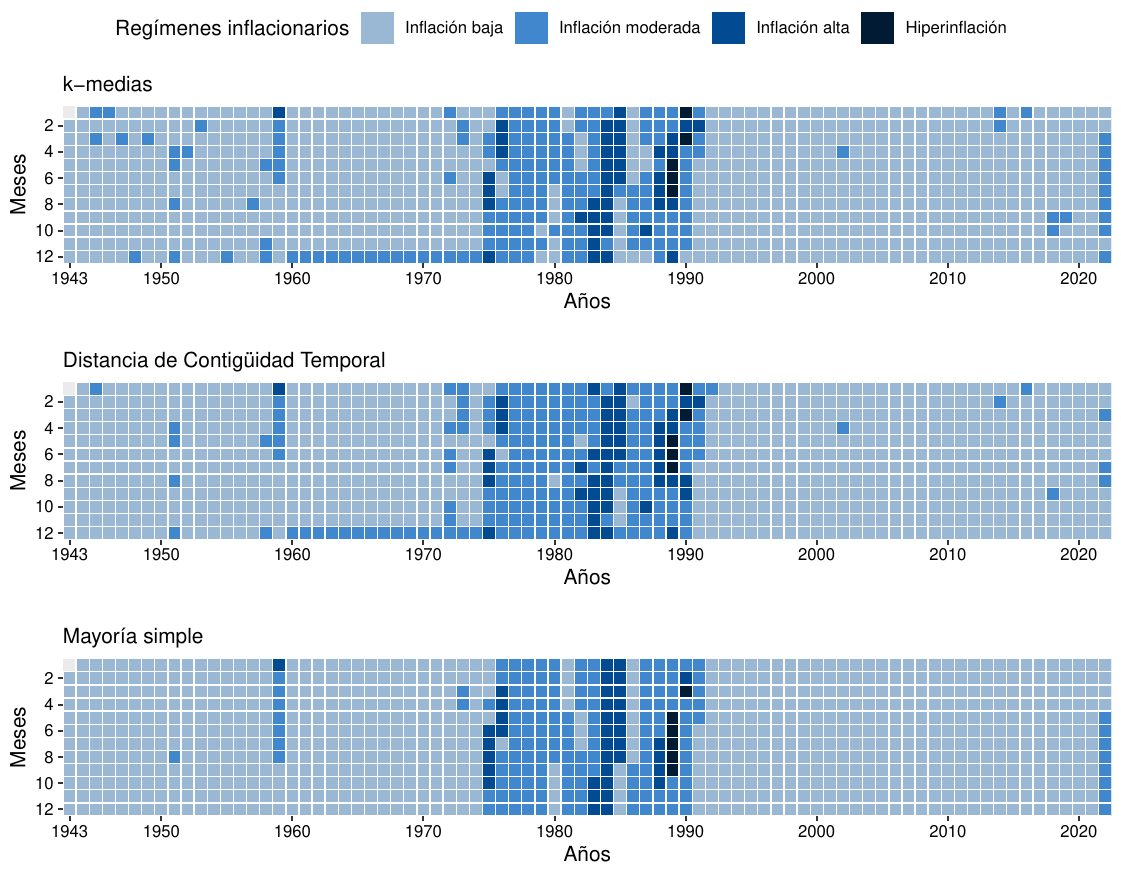}}
\vspace{-9mm}
\end{figure}

La Figura \ref{fig:comp_per} presenta gráficamente la comparación de las periodizaciones para los tres métodos. En línea con los resultados de la tabla, puede notarse como la DCT permitió suavizar la clasificación, corrigiendo las asignaciones de un grupo de observaciones en las primeras décadas de la muestra, así como asignando al régimen de inflación moderada a un conjunto de datos entre 1979 y 1987. Sin embargo, no alcanzó el resultado esperado en los meses de diciembre de la década de 1960. Por otro lado, la suavización por mayoría simple presentó un gran desempeño en dicho sentido, reduciendo casi en su totalidad los cambios abruptos de regímenes.

\subsection{Discusión}

Para evaluar los regímenes generados a la luz de la literatura previa, se realiza un ejercicio de comparación basado en el análisis desarrollado por \textcite{dabus2000}, luego extendido por \textcite{caraballoregim}, para la relación inflación-precios relativos. La Tabla \ref{tab:comp_dabus} presenta una comparación de los regímenes para toda la muestra, los cuales son comparados en función de su orden, lo que implica que el régimen de inflación baja de este trabajo se tomará como equivalente al régimen de inflación moderada de \textcite{dabus2000}. Esta clasificación presenta grandes diferencias con las expuestas en este trabajo, asignando un mayor número de observaciones a los regímenes de inflación moderada y alta, en detrimento del de inflación baja.

\begin{table}[H]
\begin{center}
\caption{Comparación de regímenes}
\label{tab:comp_dabus}
\vspace{2mm}
\footnotesize
\begin{tabular}{lccc|c}
  \hline
Regímenes & $k$-medias & DCT & MS & Dabús\\ 
  \hline
Inflación baja (moderada)		& 738 & 724 &  771 & 491 \\ 
Inflación moderada (alta) 	& 175 & 183 &  144 & 376 \\ 
Inflación alta (muy alta)		& 41  &  47 &  38 & 86 \\ 
Hiperinflación (\textit{idem}) 		& 5   &   5 &   6 & 6 \\ 
   \hline
Cambios de régimen 	 & 135   &   112 &   30 & 258\\
Cambios r/$k$-medias & -   & 58  &  92 & 294 \\
\hline
\multicolumn{5}{c}{\footnotesize Nota: Nomenclatura de \textcite{dabus2000} en paréntesis.}\\
\end{tabular}
\end{center}
\vspace{-0.5cm}
\end{table}

Con casi el doble de cambios de régimen que la estimación de $k$-medias, se puede observar cómo la clasificación de \textcite{dabus2000} es mucho menos suave en el tiempo. Bajo esta clasificación, se observa que en aproximadamente una cuarta parte de la muestra el régimen en un momento dado difiere del régimen del período anterior. 

\begin{figure}[H]
\centering
\caption{Clasificación temporal ($k$-medias, Dabús)}
\label{fig:comp_dabus}
\vspace{1mm}
\makebox[\textwidth][c]{\includegraphics[width=1.1\textwidth]{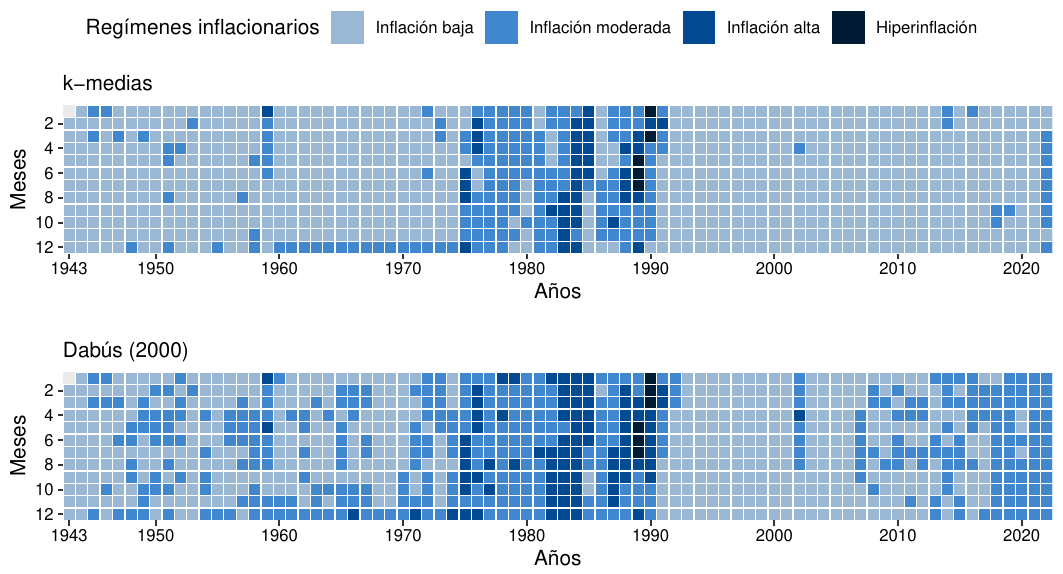}}
\vspace{-9mm}
\end{figure}

Para evaluar las clasificaciones para la relación inflación-precios relativos se emplean datos del nivel general y a nivel de capítulos de la inflación del INDEC para el período entre septiembre de 1989 y diciembre de 2006, inclusive.\footnote{ La selección de este período se debe a la disponibilidad de datos desagregados del índice de precios.} A su vez, se utilizan los regímenes estimados con la muestra completa en las secciones previas. Luego, en base a \textcite{caraballoregim} se construye la siguiente medida de variabilidad de precios relativos:
\setlength{\abovedisplayskip}{10pt}
\setlength{\belowdisplayskip}{8pt}
\begin{equation}
VPR_{t} =\frac{\sum _{i} w_{it}( \pi _{it} -\pi_{t})^{2}}{( 1+\pi_{t})^{2}}
\end{equation}

donde $w_{it}$ es la ponderación del capítulo en el índice general, $\pi_{it}$ es la tasa de inflación del capítulo $i$ y $\pi_{t}$ es la tasa de inflación del nivel general de precios.

Con estos datos, se estima un modelo lineal de la variabilidad de precios relativos en función de la tasa de inflación mensual. Sobre este modelo se evalúan la presencia de cambios de estructura para los regímenes inflacionarios desarrollados en este trabajo ($k$-medias, DCT, MS), en relación a los presentados por \textcite{dabus2000}. La Tabla \ref{tab:comp_dab_reg} presenta las estimaciones obtenidas.

\begin{table}[H]
\begin{center}
\caption{Regresiones inflación-precios relativos \label{tab:comp_dab_reg}}
\vspace{2mm}
\footnotesize
{
\def\sym#1{\ifmmode^{#1}\else\(^{#1}\)\fi}
\begin{tabular}{l*{5}{c}}
\hline
 & \textbf{Base} & \textbf{$k$-medias} & \textbf{DCT} & \textbf{MS} & \textbf{Dabús}\\
            & VPR & VPR & VPR & VPR & VPR\\
\hline
\multicolumn{6}{c}{}\\[-0.9em]

INF        &      0.0069\sym{***}&      0.0076         &      0.0076         &      0.0140         &      0.0075         \\
            &    (0.0007)         &    (0.0110)         &    (0.0124)         &    (0.0094)         &    (0.0206)         \\
[0.5em]
MOD     &                     &      0.0007         &     -0.0011         &      0.0010\sym{*}  &     -0.0001         \\
            &                     &    (0.0011)         &    (0.0008)         &    (0.0005)         &    (0.0007)         \\
[0.5em]
ALTA     &                     &      0.0006         &     -0.0036\sym{**} &      0.0045         &     -0.0007         \\
            &                     &    (0.0024)         &    (0.0013)         &    (0.0059)         &    (0.0009)         \\
[0.5em]
HIPER     &                     &      0.0033         &      0.0033         &      0.0097\sym{***}&      0.0094\sym{***}\\
            &                     &    (0.0020)         &    (0.0020)         &    (0.0011)         &    (0.0016)         \\
[0.5em]
MOD $*$ INF &                     &      0.0051         &      0.0294         &     -0.0062         &      0.0044         \\
            &                     &    (0.0152)         &    (0.0155)         &    (0.0096)         &    (0.0255)         \\
[0.5em]
ALTA $*$ INF &                     &      0.0158         &      0.0245         &      0.0000         &      0.0213         \\
            &                     &    (0.0123)         &    (0.0128)         &         (.)         &    (0.0210)         \\
[0.5em]
HIPER $*$ INF &                     &     -0.0055         &     -0.0055         &     -0.0154         &     -0.0091         \\
            &                     &    (0.0111)         &    (0.0125)         &    (0.0095)         &    (0.0206)         \\
[0.5em]
Constante      &      0.0003\sym{*}  &      0.0001         &      0.0001         &      0.0000         &      0.0001         \\
            &    (0.0001)         &    (0.0001)         &    (0.0001)         &    (0.0001)         &    (0.0001)         \\
\multicolumn{6}{c}{}\\[-1.1em]
\hline
\multicolumn{6}{c}{}\\[-1.1em]
\(R^{2}\) ajustado &       0.297         &       0.595         &       0.608         &       0.561         &       0.558         \\
\textit{AIC}&     -2057.1         &     -2167.0         &     -2173.8         &     -2150.8         &     -2148.6         \\
\textit{BIC}&     -2050.4         &     -2140.2         &     -2147.0         &     -2127.3         &     -2121.9          \\
\textit{RMSE}        &      0.0018         &      0.0014         &      0.0013         &      0.0014         &      0.0014         \\
Observaciones       &         210         &         210         &         210         &         210         &         210         \\
\multicolumn{6}{c}{}\\[-1.1em]
\hline
\multicolumn{6}{c}{}\\[-1.1em]
\multicolumn{6}{l}{\footnotesize Nota: Desvíos estándar entre paréntesis.
}\\
\multicolumn{6}{l}{\footnotesize *, **, *** indican significatividad al 5\%, 1\% y 0,1\% , respectivamente.
}\\
\end{tabular}
}
\end{center}
\vspace{-0.5cm}
\end{table}

\begin{table}[H]
\begin{center}
\footnotesize
\begin{tabular}{l*{4}{c}}
\multicolumn{5}{l}{\small Pruebas de cambio de estructura (Chow)}\\
\multicolumn{5}{c}{}\\[-1.3em]
\hline
 & \textbf{$k$-medias} & \textbf{DCT} & \textbf{MS} & \textbf{Dabús}\\
\hline
\textit{F}      &      26.50         &      28.47         &      25.92         &      21.46         \\
p-valor     &         0.000         &         0.000         &         0.000         &         0.000         \\
\hline
\end{tabular}
\end{center}
\vspace{-0.5cm}
\end{table}


Para todas las clasificaciones de regímenes se concluye en favor de la presencia de quiebres en la relación inflación-precios relativos, de acuerdo con los resultados del test de \textcite{chowtest}. Cabe destacar que, a pesar de no ser significativos individualmente, los términos de interacción resultan significativos conjuntamente para todas las clasificaciones. Estos resultados sugieren la presencia de distintas relaciones entre la inflación y los precios relativos en diferentes regímenes inflacionarios, lo cual se alinea con las investigaciones previas.

En términos comparativos, las tres clasificaciones presentadas muestran un mejor ajuste, en relación a los regímenes de \textcite{dabus2000}.\footnote{ Dado que se trata de un ejercicio de ajuste, los resultados son invariantes a la consideración de problemas de heterocedasticidad y/o autocorrelación, ya que no dependen de los errores estándar de los coeficientes.} De las tres, el procedimiento por mayoría simple exhibe relativamente el peor rendimiento. Una posible interpretación es que existe un \textit{trade-off} entre interpretabilidad y capacidad explicativa, es decir, la mayor interpretabilidad de los resultados se logra a costa de una pérdida de poder explicativo debido a los numerosos cambios realizados con respecto a la clasificación original.

\section{Conclusiones}

En este trabajo, se buscó brindar una clasificación de regímenes inflacionarios que mejorara en términos metodológicos respecto a la literatura existente, apelando a una combinación de técnicas de aprendizaje automático. Los regímenes obtenidos mediante estos métodos son específicos a la historia inflacionaria argentina y pueden considerarse, en comparación con estudios previos, como relativamente libres de sesgos por parte del investigador. Además, estos regímenes no están formados en base a una agrupación forzosa de los datos, y son independientes de las relaciones de interés sobre las cuales luego se quieran evaluar la presencia de quiebres, a diferencia de los modelos autorregresivos.

Por otra parte, se introdujeron dos procedimientos para lograr una periodización suave a lo largo del tiempo. Ambas estrategias lograron efectivamente una periodización con menores cambios de regímenes en el tiempo, facilitando la interpretación y el análisis histórico de los resultados.

Finalmente, se evaluó el poder explicativo de las clasificaciones presentadas en comparación con la proporcionada por \textcite{dabus2000} para la relación entre inflación y precios relativos. Las tres clasificaciones expuestas ($k$-medias, DCT y MS) lograron un mejor ajuste a los datos. En esto radica la relevancia del enfoque propuesto, dado que no solo posee una virtud en términos metodológicos, sino que también presenta una mayor capacidad explicativa en relación con los regímenes previos. Esto contribuye a una comprensión más profunda de la dinámica inflacionaria, que es fundamental para el desarrollo de mejores modelos de pronóstico.

Sin embargo, el enfoque metodológico expuesto presenta ciertas limitaciones. En primer lugar, las técnicas de \textit{clustering} se caracterizan por su gran sensibilidad a los datos utilizados, lo cual hace a los resultados muy inestables ante pequeñas alteraciones en la muestra empleada. A su vez, las particularidades de la temática llevaron a que no se pudiera recurrir a criterios objetivos para seleccionar la cantidad de regímenes. Por otro lado, la escasa disponibilidad de datos y la manipulación de las estadísticas oficiales de inflación condujeron a que solo se puede usar la tasa de inflación mensual como referencia, sin poder capturar otros aspectos relevantes de los regímenes.

Basados en los resultados obtenidos, se proponen explorar futuras líneas de investigación con el objetivo de mejorar y ampliar la metodología propuesta. Una de las áreas de interés sería extender la aplicación a una muestra más amplia de países, para comparar los regímenes obtenidos entre ellas. Por otro lado, resulta necesario desarrollar una técnica de \textit{clustering} temporal, que permita suavizar la clasificación en el tiempo, independientemente de la medida utilizada.

\setlength\bibitemsep{0.5\baselineskip} 
\printbibliography

\appendix

\section{Número óptimo de clusters} \label{appendix:appA}

\begin{figure}[H]
\centering
\caption{Regla del codo}
\label{fig:elbow}
\vspace{2mm}
\includegraphics[width=0.8\textwidth]{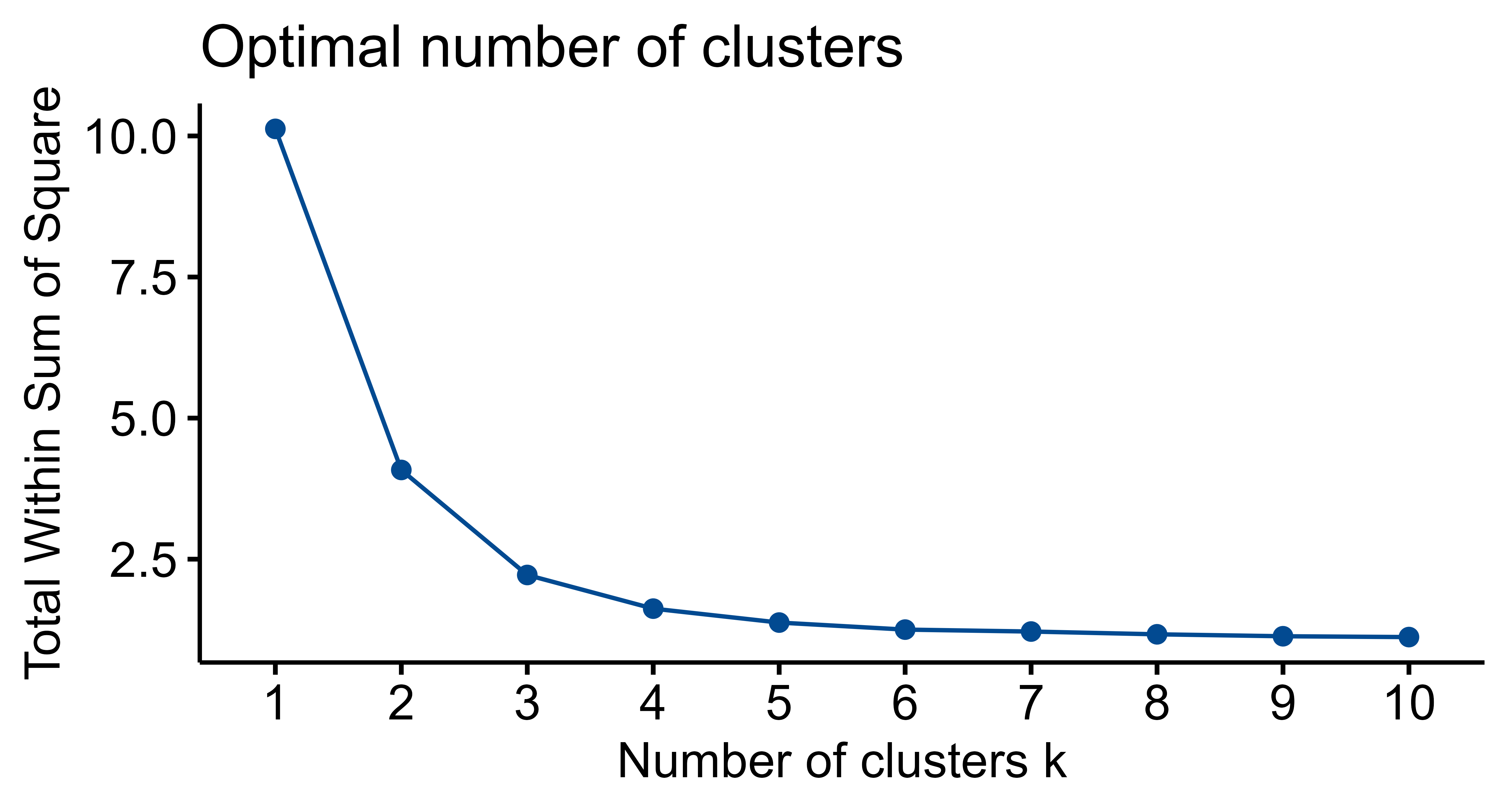}
\end{figure}

\begin{figure}[H]
\centering
\caption{Método de \textit{Silhoutte} promedio}
\label{fig:sil width}
\vspace{2mm}
\includegraphics[width=0.8\textwidth]{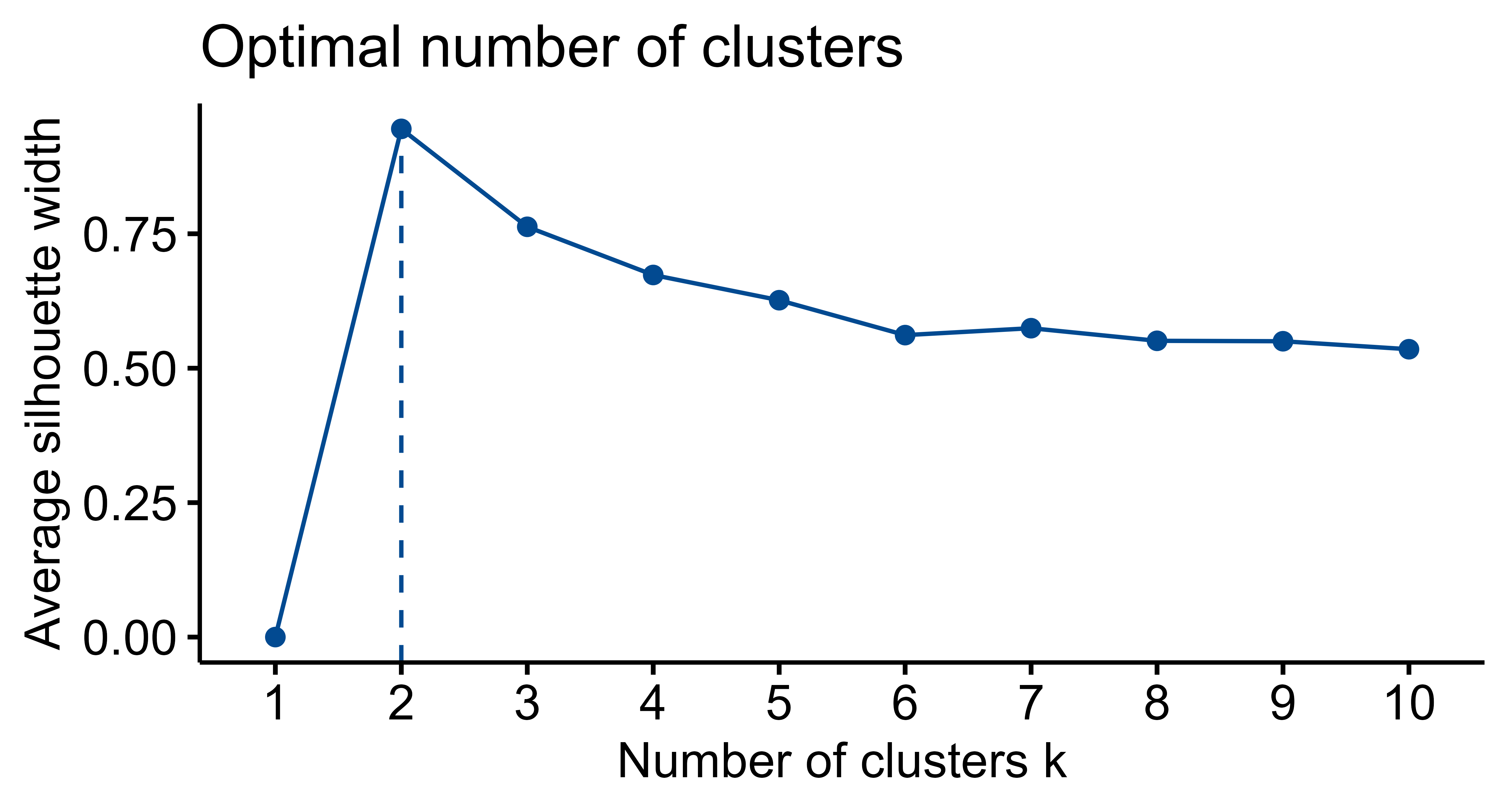}
\end{figure}

\begin{figure}[H]
\centering
\caption{\textit{Gap statistic}}
\label{fig:gap stat}
\vspace{2mm}
\includegraphics[width=0.8\textwidth]{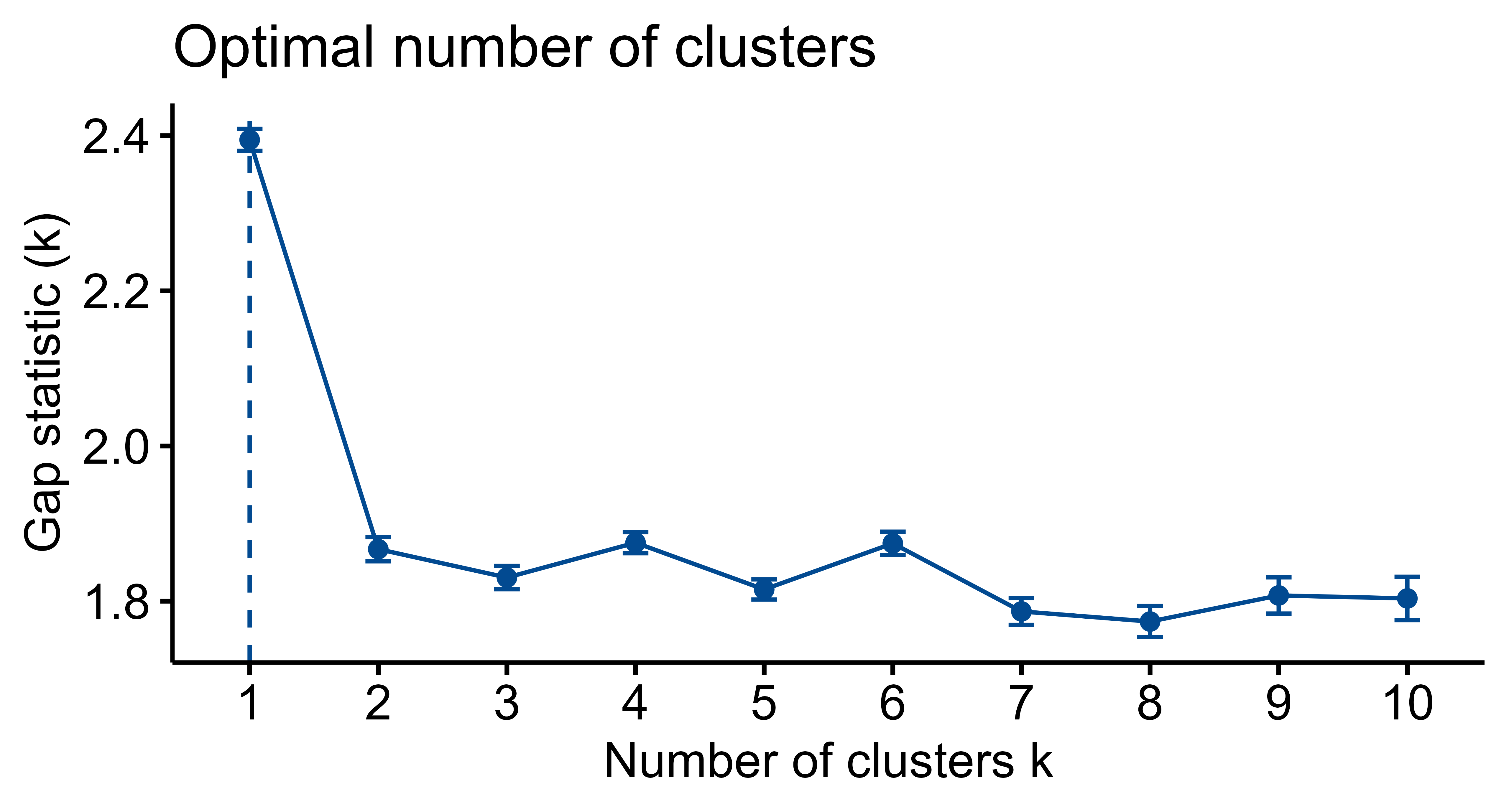}
\end{figure}

\end{document}